\documentclass[fleqn,10pt]{wlscirep}
\usepackage[utf8]{inputenc}
\usepackage[T1]{fontenc}
\usepackage{graphicx}
\usepackage{amsmath}
\usepackage{amssymb}
\usepackage{physics}
\usepackage{hyperref}
\title{A fibered laser system for the MIGA large scale atom interferometer}
\author[1]{D. O. Sabulsky}
\author[1,2]{J. Junca}
\author[1]{G. Lef\`{e}vre}
\author[1]{X. Zou}
\author[1]{A. Bertoldi}
\author[1]{B. Battelier}
\author[3]{M. Prevedelli}
\author[2]{G. Stern}
\author[2]{J. Santoire}
\author[4]{Q. Beaufils}
\author[4]{R. Geiger}
\author[4]{A. Landragin}
\author[2]{B. Desruelle}
\author[1]{P. Bouyer}
\author[1,$\dagger$]{B. Canuel}
\affil[1]{LP2N, Laboratoire Photonique, Numérique et Nanosciences, Université Bordeaux-IOGS-CNRS:UMR 5298, rue F. Mitterrand, F-33400 Talence, France}
\affil[2]{MUQUANS, Institut d’Optique d’Aquitaine, rue F. Mitterrand, 33400 Talence, France}
\affil[3]{
Dipartimento di Fisica e Astronomia, Universit{\`a} di Bologna,
Via Berti-Pichat 6/2, I-40126 Bologna, Italy}
\affil[4]{LNE-SYRTE, Observatoire de Paris, Universit\'e PSL, CNRS, Sorbonne Universit\'e, 61 avenue de l’Observatoire, 75014 Paris, France}
\affil[$\dagger$]{benjamin.canuel@institutoptique.fr}


\begin{abstract}
We describe the realization and characterization of a compact, autonomous fiber laser system that produces the optical frequencies required for laser cooling, trapping, manipulation, and detection of $^{87}$Rb atoms - a typical atomic species for emerging quantum technologies.
This device, a customized laser system from the Muquans company, is designed for use in the challenging operating environment of the Laboratoire Souterrain \`{a} Bas Bruit (LSBB) in France, where a new large scale atom interferometer is being constructed underground - the MIGA antenna.
The mobile bench comprises four frequency-agile C-band Telecom diode lasers that are frequency doubled to 780 nm after passing through high-power fiber amplifiers. 
The first laser is frequency stabilized on a saturated absorption signal via lock-in amplification, which serves as an optical frequency reference for the other three lasers via optical phase-locked loops. 
Power and polarization stability are maintained through a series of custom, flexible micro-optic splitter/combiners that contain polarization optics, acousto-optic modulators, and shutters.
Here, we show how the laser system is designed, showcasing qualities such as reliability, stability, remote control, and flexibility, while maintaining the qualities of laboratory equipment.
We characterize the laser system by measuring the power, polarization, and frequency stability.
We conclude with a demonstration using a cold atom source from the MIGA project and show that this laser system fulfills all requirements for the realization of the antenna. 
\end{abstract}
\begin{document}
\flushbottom
\maketitle
\section*{Introduction}
\par Large research infrastructures require the production of customized, well-engineered equipment to meet demanding scientific requirements, and for use in environmental conditions that often goes beyond standard laboratory ones. 
Such requirements stand at the frontier of industrial and laboratory grade qualities: these devices need to be stable, be remote controlled, and have the possibility of mass-production while maintaining flexibility and high performance. 
The Matter wave-laser based Interferometer Gravitation Antenna (MIGA) \cite{Canuel2018}, is a large scientific instrument presently under construction at the Laboratoire Souterrain \`{a} Bas Bruit (LSBB) \cite{Bettini2012} in Rustrel, France. 
 MIGA is based on an array of atom interferometers simultaneously manipulated by a resonant light field circulating in a 150 m optical cavity, creating a state-of-the-art gradiometer. This antenna will be the first underground infrastructure dedicated for matter-wave experimentation; it will pave the way toward the realization of a new class of gravitational wave detectors in the infrasound domain \cite{Chaibi2016} and enable to study gravity gradient fluctuations at low frequency \cite{Junca19}.
This large scale antenna requires dedicated, automated instrumentation subsystems to make the infrastructure cost effective and simple to maintain while accomplishing scientific goals. 
One of the most challenging of these subsystems is the laser apparatus to control the numerous atomic sources within the antenna.
This devices must comply with environmental and technical requirements that differs significantly from those of typical laboratory applications:
the environment within the underground LSBB laboratory is challenging, with galleries located at a depth up to 500 m below a karstic massif, where the system must operate remotely and stably without frequent maintenance.
These constrains necessitate a fiber optical solution: telecommunications C-band equipment is reliable, well-tested, and has demonstrated ability to work under non-laboratory conditions; critically, this solution does not need constant maintenance.
The laser system must also be able to meet the requirements of a complex, modern cold atom system; it needs to cool atomic clouds launched from the atomic source to a 3D kinetic temperature in the range of a few $\mu$K, in a single magnetic substate, and in a narrow range of velocities along the axis of the laser interferometer. 
It is also required to provide all laser frequencies needed for a detection system based on fluorescence.
This laser system system, based on reliable telecommunications technology, has applications toward mobile, field deployable quantum technologies.
\par Pioneering experiments in matter-wave interferometry \cite{Carnal1991, Keith1991, Riehle1991} demonstrating inertial sensitivity led to significant interest for precision measurements using atomic beamlines. 
Development of laser-cooling schemes \cite{Chu1998, Cohen1998, Phillips1998, Metcalf1999} during the last thirty years have produced a series of methods to produce small 3D kinetic temperatures in dilute atomic gases.
This led to atom interferometry with cold atoms, which has a track record of accuracy and precision spanning the development of laser cooling. 
This technique has been used to measure gravity \cite{Kasevich1991, Peters1999, Karcher_2018, Freier2016, Wang2018, Bidel2018}, gravity gradients \cite{McGuirk2002, Sorrentino2014}, rotations \cite{Gustavson1997, Gauguet2009, dutta, dutta:tel-01332562, Savoie2018ScAdv}, measurements of the gravitational constant G \cite{Fixler2007, Rosi2014}, determination of the fine structure constant coupled with tests of quantum electrodynamics \cite{Bouchendira2011, Parker2018}, studies of the interface between gravity and quantum mechanics \cite{Asenbaum2017, Geiger2018}, General Relativity (GR) tests such as the searches for violations of the Weak Equivalence Principle \cite{Dimopoulos2007, Bonnin2013, Tarallo2014, Schlippert2014,Barrett2015, Barrett2016, Duan2016}, tests of local Lorentz invariance of post-Newtonian gravity \cite{Muller2008_5}, experiments in microgravity \cite{condon2019alloptical, JPLCAL, Becker_2018}, tests of dark energy \cite{sabulsky, hamilton849}, and recoil associated with blackbody radiation \cite{Haslinger_2017}.
Based on this record of precision measurements, cold atom interferometry gained traction as a candidate system for observing Gravitational Waves (GWs) at low frequency \cite{Borde1983, Tino2007, Dimopoulos2008, Yu2010, Graham2013PRL, Zhan2019, canuel2019elgar}.
The schemes developed in these experiments form a core technology under the auspice of emerging quantum technologies, as a new type of deployable sensor - atom interferometry is being applied to inertial navigation \cite{Canuel2006, Geiger2011, PhysRevApplied.10.034030}, space and satellite missions \cite{Aguilera2014, Trimeche2019,tino2019sage, bertoldi2019aedge}, mineral prospecting \cite{Romaides2001, metje}, and civil engineering \cite{Hinton2017}.
These applications require, and have led to, the development of reliable, stable, mobile, and compact laser systems to control, manipulate, and measure the atomic samples under varying environmental conditions; these devices have increasingly relied upon frequency doubling of stable C-band telecommunications lasers and the related developments to make such lasers rugged and reliable \cite{Thompson:03, Masuda:07, Lienhart2007, mnoret2011, Zhang_2012, leveque_2014, Matthey:15, WANG201682, THERON2017152, caldani2019prototype}.
\par Muquans \cite{Muquans}, the first company to bring to market a completely fibered laser apparatus dedicated to cold atom production and atom interferometry \cite{Menoret2018}, has been selected as the commercial entity to realize the automated fiber laser systems for the MIGA antenna.
In this report, we present the first realization and characterization of such a system, designed for the MIGA project. 
We demonstrate that this device fulfills its application requirements from measurements on a prototype atomic source of the antenna.
First, we present a series of schematic diagrams of the laser system and discuss its internal subsystems.
We then report on the performance of the device, measuring the stability of polarization extinction ratio (PER), output intensity, and frequency.
Finally, we test the laser by applying it to the cooling, launching, and preparing of clouds of dilute $^{87}$Rb, which requires all the functionalities of the laser system working in concert. 
\section*{The laser system for MIGA}
\begin{figure}[hbt]
\centering
\includegraphics[width=0.85\linewidth]{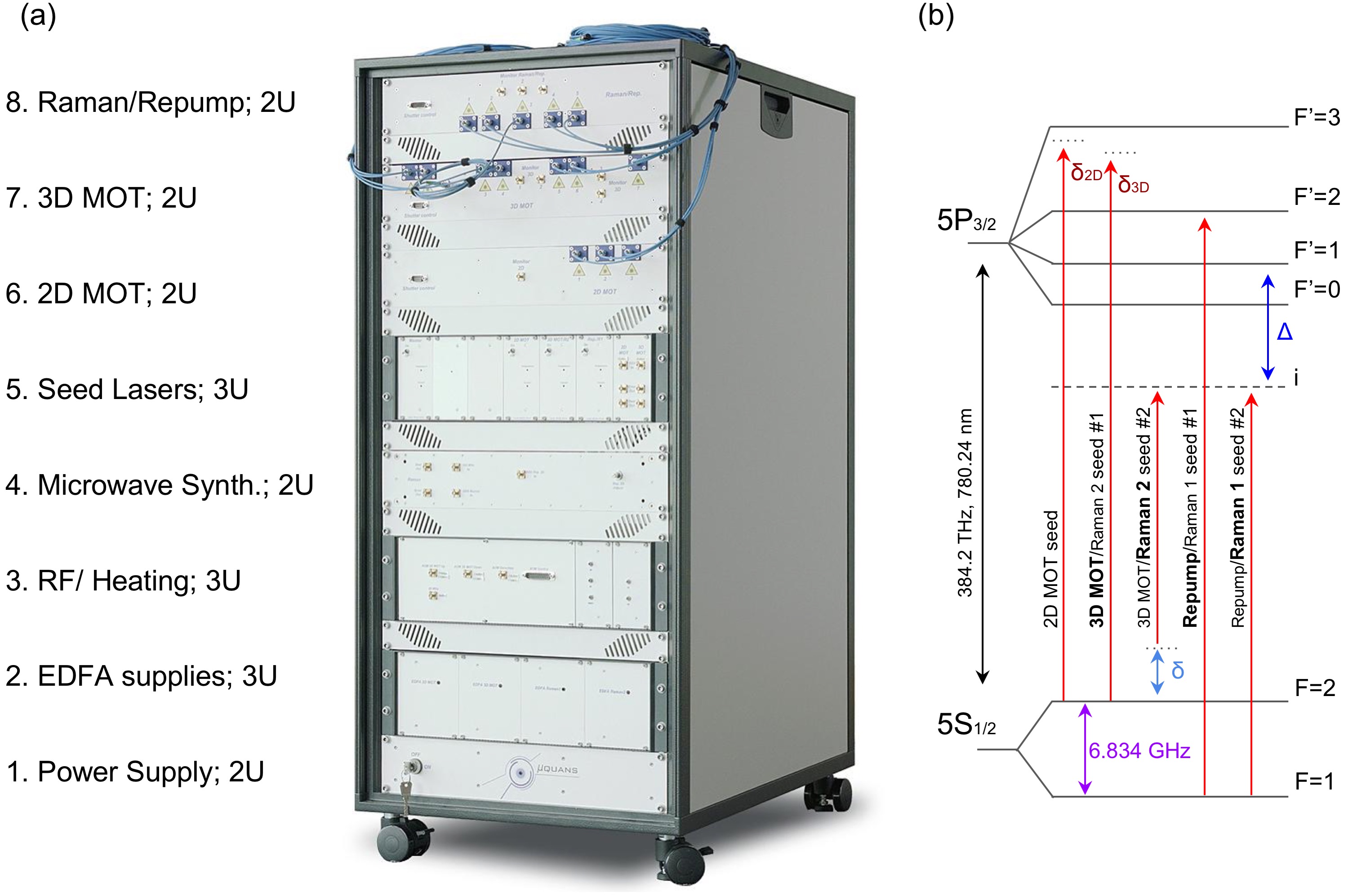}
\caption{
(a) The automated laser system inside of a typical mobile 19" rack, without thermal/condensation shroud.
There are eight 2U and 3U rack units with the following functions, from bottom to top: the power supply converting 230 V AC to various DC voltages (1), the high power EDFA controllers (2), a combined unit of RF components and temperature servos (3), a dedicated microwave frequency unit (4), cubbyholes for the lasers and associated electronics (5), the 2D MOT micro-optic module with integrated PPLN crystal and shutter (6), 3D MOT micro-optic module with integrated AOMs, PPLN crystal, and shutters (7), and the Raman/Repump micro-optic modules with integrated AOMs, PPLN crystals, and shutters (8). 
(b) Seed lasers locking frequencies relative to $^{87}$Rb D$_2$ transitions.
Cooling inside the 3D MOT and driving Raman transitions employ the same pair of seeds 3D MOT/Raman 2 and Repump/Raman 1, by changing their frequency set-point. We label \#1 and \#2 as the corresponding set-points configurations.
Configuration \#1:
the 3D MOT/Raman 2 seed is red-detuned $\delta_{\text{3D}}$ from the $F = 2$ to $F' = 3$ transition, and the Repump/Raman 1 seed is tuned to be resonant to the $F = 1$ to $F' = 2$ transition.
Configuration \#2:
the 3D MOT/Raman 2 seed is red-detuned by $\Delta$ (GHz) from the $F = 2$ to $F' = 1$ transition, and the Repump/Raman 1 seed is red-detuned by $\Delta$ from the $F = 1$ to $F' = 1$ transition.
We scan the frequency of the Raman transition, with respect to the hyperfine splitting 6.834 GHz, by applying $\delta$ (kHz to MHz).
The 2D MOT seed laser is red-detuned $\delta_{\text{2D}}<\delta_{\text{3D}}$ from the $F = 2$ to $F' = 3$ transition. For this seed, the repump frequency is addressed for the 2D MOT by frequency modulation with an EOM.
\label{fig:laserimage}} 
\end{figure}

The core technologies of the autonomous fiber laser system are a result of advances in C-band telecommunications devices. 
All lasers in the system comply with Telcordia qualification GR-468 \cite{tele}.
These external cavity diode lasers (ECDLs) are routinely offered with linewidths $\leq$ 10~kHz  at 1~s and optical powers of $\sim$ 10~mW in a compact butterfly package.
We use lasers tuned to 1560~nm, which, when frequency doubled through a periodically-poled Lithium Niobate (PPLN) crystal to 780~nm, are capable of cooling, trapping, and manipulating atomic Rubidium. 
Our application requires significant power after frequency doubling; C-band telecommunications technology offers high power, low noise, fiber-optic solutions for applications requiring powers in excess of 100 mW  in the form of Erbium-doped fiber amplifiers (EDFAs). 
Combined with significant improvements to and cost reductions of polarization-maintaining, portable, stable micro-optic benches, complex configurations are easily realized \cite{kylia}.
The MIGA project laser system is a combination of these advances - a fibered, compact, autonomous, plug-and-play racked system, see (a) in Figure~\ref{fig:laserimage}. 
It can be completely remote controlled through a terminal interface via Ethernet. 
This connection allows communication with the laser as well as monitoring of the various subsystems.
The system is contained within a climate controlled, mobile rack.
The climate control of the rack creates a mobile laboratory-standard environment - this allows the laser system to operate at the LSBB unimpeded by local temperature and humidity.
The laser system has the dimensions 117~cm $\times$ 80~cm $\times$ 55~cm, requires 100 - 240~V$_{\text{AC}}$ at 50-60~Hz for 300~W max power, and weighs $\simeq$150~kg. 
\subsection*{Design Principle}
The system is based on four fibered ECDLs primary oscillators that are frequency stabilized.
The first is a reference laser, which is used by the other three oscillators as a fixed frequency set-point. 
These other three oscillators seed EDFAs - we refer to these primary oscillators as seed lasers.
They are labelled as the 2D MOT seed, the 3D MOT/Raman 2 seed, and the Repump/Raman 1 seed, and are frequency stabilized to address various transitions in the $^{87}$Rb D$_2$ spectrum \cite{steck2001rubidium} enabling the laser cooling of $^{87}$Rb atoms in a 2D-3D MOT configuration and to realize Raman spectroscopy (see (b) of Figure~\ref{fig:laserimage}).
The aforementioned seed lasers are frequency agile, in that their frequency can be shifted by 1 GHz in less than $1$ ms, which is useful for reducing the number of seed lasers required to run the system but also to operate in an interleaved measurement configuration \cite{PhysRevLett.111.170802, Savoie2018ScAdv}.
In this way,  we can realize both cooling in the 3D MOT and Raman transitions with the same pair of seeds 3D MOT/Raman 2 and Repump/Raman 1, by changing their frequency set-point.
To control these lasers and the system peripherals, we designed the control inputs of the system to match a home made control system \cite{ctrlsys}; we cascade the control systems, enabling the control of many lasers through a single remote interface. 
The tests reported in the present paper were carried out with this dedicated control system but we emphasis that any computer control system that has analog and digital outputs can operate the laser. 
\subsubsection*{Laser Architecture}
\begin{figure}
\centering
\includegraphics[width=1\linewidth]{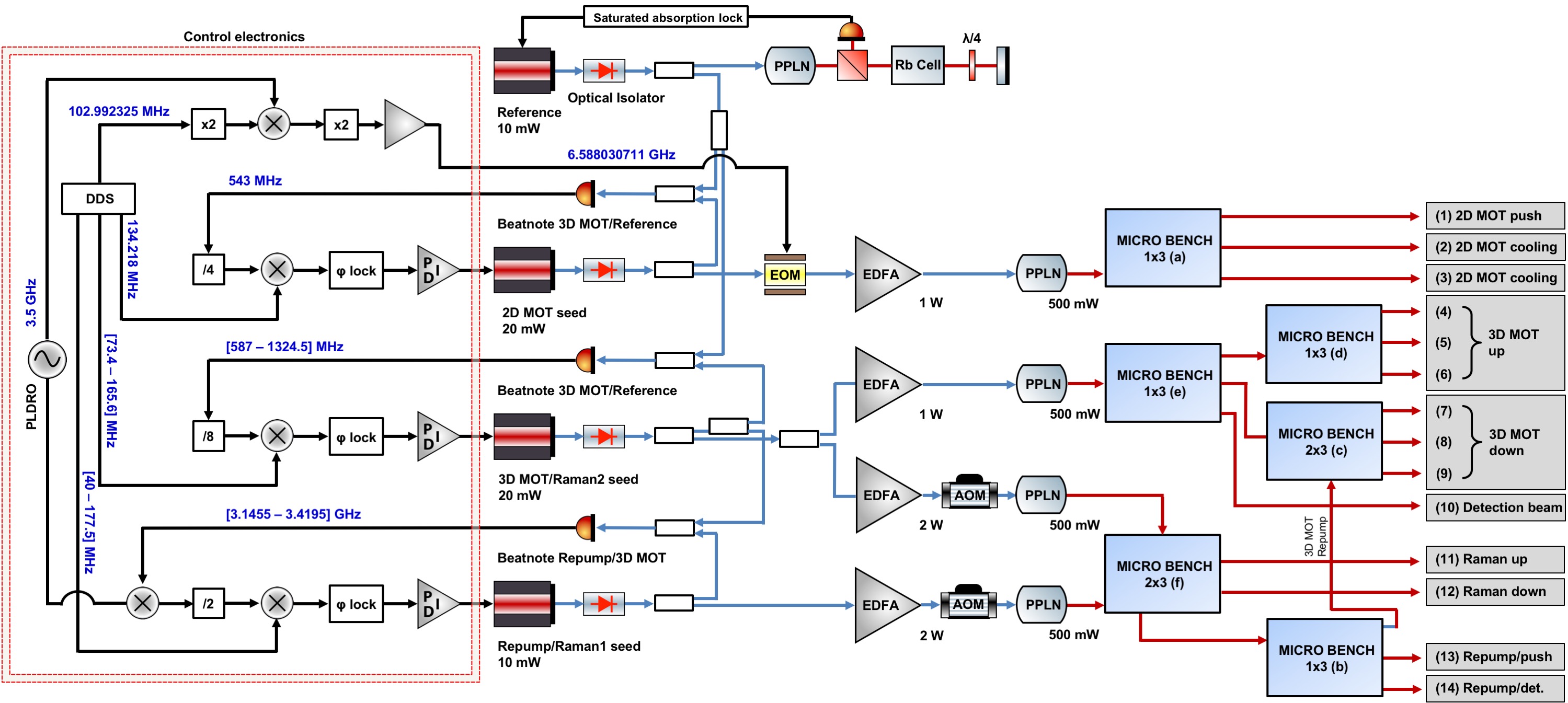}
\caption{Schematic of the laser system. 
Each primary oscillator passes light through a fibered optical isolator and into a combination of fused fiber splitter/combiners to create beatnotes.
The three primary oscillators that are frequency stabilized via beatnotes seed four EDFAs, which in turn seed four PPLN crystals.
The Raman outputs are controlled using fibered AOMs at 1560 nm, after the amplifiers but before the PPLN crystals. 
After the PPLN crystals, the light is sent to various micro-optic modules to split, combine, and output the light via PM optical fibers.
AOMs contained within the 1 $\times$ 3 micro-optic bench, (e) in Figure~\ref{fig:micromodule}, provide control of the 3D MOT fibers. 
To create a frequency reference for the RF control of the laser system, we employ a PLDRO to generate a 3.5 GHz signal for mixing with the high frequency beatnote in the system - the beat between 3D MOT/Raman 2 and Repump/Raman 1; we exert control over the Repump/Raman 1 laser via mixing with a DDS. 
A part of the power of the PLDRO is mixed with a DDS control frequency and doubled to drive the EOM in the 2D MOT chain that addresses the repump transition. 
The other beat notes are controlled similarly but do not require a high frequency reference like the PLDRO - we divide down the various beatnotes and mix directly with DDS signals. 
The PLDRO and DDS are phase locked to a 100 MHz input signal.
RF and microwave connections are in black, 1560 nm PM fibers are in blue, and 780 nm PM fibers are in red. 
\label{fig:laserspewpew}}
\end{figure}

The complete laser architecture can be seen in Figure \ref{fig:laserspewpew} and is composed of four lasers, four amplifers, four PPLN doubling crystals, one Electro-Optic Modulator (EOM), five Acousto-Optic Modulators (AOMs), and six different micro-optic benches. 
One of the four primary oscillator, the reference laser, is frequency stabilized to the largest spectroscopic cross-over feature in the $^{85}$Rb spectrum by lock-in amplification on saturated absorption spectroscopy.
This spectroscopy cell is around a few cm in length and is encased in a magnetic shield. 
The other three are frequency stabilized via beat notes to this reference with optical phase-lock loops (OPLLs); these oscillators seed EDFAs, which in turn seed fibered PPLN crystals.
The linewidths of all lasers were measured to be less than 40 kHz at 1 second, at 780 nm.
\par The laser system is clocked with a 100 MHz input - this is used to synchronize the DDS that controls the various laser beatnotes and a phase-locked dielectric resonant oscillator (PLDRO).
We create microwaves near to the $^{87}$Rb hyperfine frequency by mixing controllable DDS signals with the 3.5 GHz PLDRO and then employing multipliers.
Frequencies in the range of 70 MHz to are generated directly by DDS and are mixed and multiplied - high frequency beatnotes are fed into the OPLLs where they are divided down and then mixed with the control frequencies. 
The lasers can be frequency modulated, by their servos, up to 200 kHz using their current; the servos have a slow path that feeds back on the diode temperature up to $\simeq0.5$ Hz. 
\par The 2D MOT seed passes light through an in-line electro-optic modulator (EOM), to illuminate the $F=1$ manifold, before undergoing amplification and frequency-doubling; see (b) in Figure~\ref{fig:laserimage} and Figure~\ref{fig:laserspewpew}.
The 3D MOT requires the two seeds, 3D MOT/Raman 2 and Repump/Raman 1, the former for the cooling transition and the latter to pump atoms out of the dark manifold $F=1$. 
By applying a frequency jump, the seed lasers driving the 3D MOT can also drive Raman transitions. 
There are four EDFAs in the system - one for the 2D MOT seed, one for the Repump/Raman 1 seed, and two for the 3D MOT/Raman 2 seed; this last laser has its amplified output split among multiple micro-optic modules. 
Each amplifier seeds a fiber coupled PPLN crystal, doubling the optical frequency from 1560 nm to 780 nm; the PPLN conversion efficiencies range from $40\%/\text{W}$ to $60\%/\text{W}$ depending upon the input power at 1560 nm -we typically use 1 to 2 W of power at 1560 nm.
To maximize this efficiency, we use an automated scan that varies the temperature of the crystal to find and set the phase-matching temperature.
The output of the PPLN crystals injects custom built micro-optic splitter/combiner modules, see Figures~\ref{fig:laserspewpew} and~\ref{fig:micromodule}.
The Raman beams pass through AOMs before frequency doubling while the cooling transition frequencies pass through integrated AOMs within the micro-optic modules after doubling; for this reason, the Raman EDFAs have a higher output power, to make up for significant insertion loss through the fiber AOMs.
We employ 6 micro-optic modules in this laser system to distribute the laser frequencies into 14 polarization-maintaining (PM) fiber optic outputs. 
There are three outputs for the 2D MOT; for the horizontal, vertical, and push beams (see outputs (1) to (3) on Figure~\ref{fig:micromodule}).
There are six outputs for the 3D MOT (see (4) to (9)), two outputs for Raman beams (see (11) and (12)), two outputs for the detection frequencies (see (13) and (14)), and one output for a beam used to clear the $F=1$ manifold of the $5S_{1/2}$ ground state (see (10)).
\subsubsection*{Micro-optic modules}
\begin{figure}[htbp]
\centering
\includegraphics[width=1\linewidth]{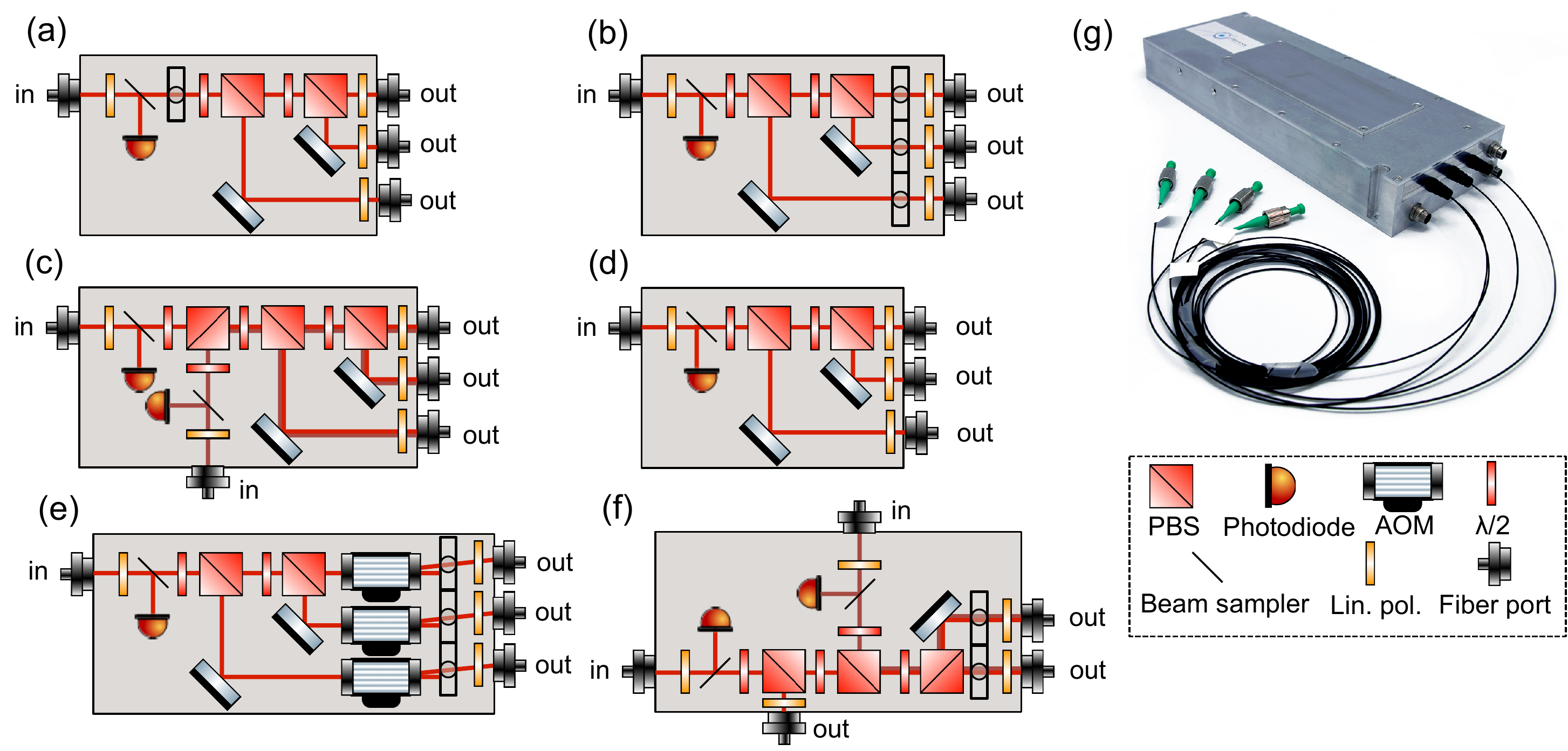}
\caption{Schematic of all micro-optic splitter/combiners employed in the laser system. 
(a) 1 $\times$ 3 micro-optic splitter with an internal shutter before splitting, to shutter all beams simultaneously.
This is used to split the 2D MOT light.
(b) 1 $\times$ 3 micro-optic splitter with internal shutters after splitting, to shutter all beams separately. 
This is used to split light from Repump/Raman 1 when tuned to the repump transition.
(c) 2 $\times$ 3 micro-optic splitter/combiner for combining two beams together, in the same polarization, into a PM fiber.
This device is for the bottom three beams of the 3D MOT, with the repump frequency combined into the beams.
(d) 1 $\times$ 3 micro-optic splitter.
This variant is used to split a beam for the upper three beams of the 3D MOT.
(e) 1 $\times$ 3 micro-optic splitter with AOMs and internal shutters after splitting, for amplitude and frequency control of each beam.
This is the device that gives control over the 3D MOT beams and detection. 
(f) 2 $\times$ 3 micro-optic splitter/combiner with shutters. 
This variant serves two purposes, to split/combine the Raman frequencies together into two fibers as well as manage a modest pick off to send repump light to other splitter/combiners, namely (b) and (c).
(g) Photo of a hermetically sealed, configuration (c) micro-optic module. 
Bottom right; a key for optical components.  
\label{fig:micromodule}} 
\end{figure}
After frequency doubling, the output of the PPLN crystals has to be distributed among a number of PM optical fibers with differing amounts of power. 
This can be done using fused fiber fiber optic beam splitter/combiners or discrete optical systems contained within ultra-stable modules. 
At 1560 nm, we employ fused fiber splitters for making beatnotes, seeding amplifiers, and withinin the amplifiers themselves (see Figure~\ref{fig:laserspewpew}), as these devices have a proven track record of reliability, stability, and ruggedness in telecom and space applications at this frequency. 
At 780 nm, polarization maintenance of fused fiber devices is less reliable, especially in challenging environmental conditions.
After frequency doubling, we  employ miniaturized free-space polarization optics to split and combine laser light at 780 nm; these micro-optic devices, manufactured by Kylia\cite{kylia} for Muquans, preserve alignment and polarization in compact, hermetically sealed modules, see Figure~\ref{fig:micromodule}. 
This option is flexible because we can easily change power outputs, and it allows an easy translation from complex optical bench setups to miniaturized, mobile bench designs.
The ability to tune the outputs is a major advantage allowing the fine-tuning of each laser system for all atomic sources.
The modules display stable thermal characteristics, which minimizes the effects of thermal drift on polarization and power stability. 
Every module has internal photodiodes to monitor the input light, temperature sensors to correlate output power drifts, and a linear polarizer in front of every input and output to ensure maximal stability.
Typical modules weigh 1 - 2 kg with dimensions $\sim 100~\text{mm}~\times \sim 250~\text{mm}~\times \sim 50~\text{mm}$.
In table \ref{tab:one} we present measurements of the performance obtained for each fiber output of all splitters: the maximum output power, the minimum polarization extinction ratio over three hours, and the RMS power noise over 70 hours.
The optical layout of the micro-optic modules is translated directly from optical table configurations.
\par The 2D MOT module can be seen on (a) of Figure~\ref{fig:micromodule}.
The frequency doubled output of the PPLN crystal is injected into the module input, first going through a linear polarizer and then a small beam sampler, for monitoring. 
After passing through a blade-type shutter, the beam power is divided among the three outputs using half-wave plates and polarizing beam splitting (PBS) cubes.
\begin{table}[ht]
\centering
\resizebox{\linewidth}{!}{\begin{tabular}{|l|l|l|l|l|l|l|l|l|l|l|l|l|l|l|}
\hline
Seed & \multicolumn{3}{c|}{2D MOT} & \multicolumn{7}{c|}{3D MOT/Raman 2} & \multicolumn{2}{c|}{3D MOT/Raman 2 and Repump/Raman 1} & \multicolumn{2}{c|}{Repump/Raman 1} \\
\hline
Fiber Output $\#$ & (1) & (2) & (3) & (4) & (5) & (6) & (7) & (8) & (9) & (10) & (11) & (12) & (13) & (14)  \\
\hline
Power (mW) & 3.4 & 184 & 183 & 40 & 38 & 37 & 39 & 38 & 36 & 43 & 443 (183 + 260) & 445 (183 + 262) & 13 & 5 \\
\hline
PER (dB) & 20 & 21 & 25 & 28 & 25 & 24 & 22 & 20 & 20 & 30 & (25 / 25) & (32 / 36) & 22 & 26  \\
\hline
RMS ($\%$) & <1 & <1.4 & <2.1 & <1 & <1.3 & <1 & <1 & <1 & <1 & <1.5 & <1 / <1.3 & <1 / <1.7 & <1.5 & <1.7 \\
\hline
\end{tabular}}
\caption{Maximum output power (mW), PER (dB) over three hours, and the RMS power stability over 70 hours of the 14 fiber outputs of the system. 
Fiber numbering matches the output labels in Figure \ref{fig:laserspewpew}.
Fibers (1) to (3) are for the 2D MOT.
Fibers (4) to (9) are for the 3D MOT. 
Not accounted for here, there is an additional 19 mW of repump light evenly distributed among (7), (8), and (9); the PER of this fiber is 21 dB.
Fiber (10) is for detection and clearing the $F=2$ manifold. 
Fiber (11) and (12) are for the Raman transitions required for state and velocity selection; the power from the amplifiers is separately shown as are the PERs. 
Fiber (13) is for clearing the $F=1$ manifold. 
Fiber (14) is for detection of atoms in the $F=1$ manifold. 
\label{tab:one}}
\end{table}
\par The 3D MOT light is divided into six outputs through three micro-optic modules, see Figure~\ref{fig:laserspewpew} as well as (c), (d), and (e) of Figure~\ref{fig:micromodule}. 
The 780 nm light from the PPLN enters the micro-optic module (e) through a linear polarizer and beam sampler for monitoring. 
The light, after power splitting through the polarization optics, is sent through single-pass AOMs, shutters, and is re-injected into fiber optic to be sent to the next splitting modules (c) and (d). 
These two modules, used to create the 3D MOT beams, are similar to the 2D MOT module - the only difference being that the module (c) for the bottom three beams of the 3D MOT has a second injected frequency, the repump light.
\par The Raman light is directly injected into the module from two PPLN crystals, see Figure~\ref{fig:laserspewpew} as well as (f) of Figure~\ref{fig:micromodule}.
We opt to use AOMs at 1560 nm because they provide a higher extinction ratio, which is preferable for the application of driving Raman transitions, despite the significant optical insertion loss.
The module (f) combines, in the same polarization, the two frequencies required for driving a Raman transition into two optical fibers; Raman 1 and Raman 2 are combined into a single beam and are split between two fibers, Raman up (11) and Raman down (12). 
A modest beam pick off from Raman 1 is sent from module (f) to module (b). This module creates the repump light sent to the 3D MOT module (c) but also the outputs (13) and (14) used for preparation and detection.
\par All power outputs are variable; they can be modified by simply opening the module and rotating the waveplates.
The alignment and stability are maintained, despite the lasers being shipped various times and moved around their host facilities.
The ability to finely tune the modules is necessary for pairing the laser to an individual atomic source, in order to meet sensor head performance requirements.
\subsubsection*{Laser frequency stabilization}
Frequency stabilization is crucial to laser cooling, trapping, and manipulation of neutral atoms.
The required stability is small by comparison to the linewidth of the $\ket{5^{2}S_{1/2}}\rightarrow \ket{5^{2}P_{3/2}}$ transition  in $^{87}$Rb ($6.07$ MHz). 
Frequency error of the 2D MOT seed and 3D MOT/Raman 2 seed directly corresponds to quadratic changes in the spring constant of the respective traps; frequency error for the preparation or detection leads to incorrect velocity selection or error in atom number counting.
To this end, the reference must be stable at the hundreds of kHz level at long times and this frequency stability must be passed on to the other seed lasers. 
The reference laser signal is split and mixed with samples of the 2D MOT seed and the 3D MOT/Raman 2 seed using a series of fused fiber splitters, see Figure~\ref{fig:laserspewpew}. 
These beat notes are used to servo the frequencies of these seed lasers via OPLLs  \cite{Santarelli1994,Schmidt2010}.
The fourth laser, Repump/Raman 1 seed, is phase-locked to the 3D MOT/Raman 2 seed.
In this way, all seed lasers have similar stability to the reference laser. 
\par The frequency chains used to control these lasers are presented on the left of Figure \ref{fig:laserspewpew}.
We exert control over the laser frequencies using DDS in combination with a series of RF multipliers, dividers, and mixers. 
In this way, we can use RF frequencies $\leq200$ MHz to control beat notes up to 7 GHz with precision and accuracy to the sub-Hz level. 
For direct manipulation of GHz frequencies, we mix in a 3.5 GHz signal from a PLDRO; it is phase-locked to an external 100 MHz signal provided by the user, typically an ovenized quartz oscillator or atomic vapor clock, itself referenced to a disciplined GPS signal. 
These signals are converted into error signals and fed into a phase-lock, creating a closed-loop frequency stabilization with an agile setpoint that the user controls.
\subsubsection*{Remote control and required system inputs}
The laser system requires a series of user inputs in the form of TTLs, control voltages, and sinusoidal waveforms in the MHz range to function. 
We employ a custom, application specific control system designed within the MIGA project\cite{ctrlsys}.
The user communicates with the laser system via ssh or telnet connection through terminal from any computer. 
The laser system has a series of internal computers for acquiring diagnostic data, controlling the amplifiers and PPLNs, and frequency stabilizing the lasers. 
The diagnostic information about the laser system includes temperature of each rack unit and critical components within each of them, setpoints for currents and voltages, live lock-loop parameters, and photodiode monitoring at 1560 nm and 780 nm from the seeds to the amplifiers and after the PPLN crystals in each micro-optic module.
These diagnostics are sampled at 1 Hz for remote monitoring, but can be increased up to 100 Hz; analog connections are available for higher bandwidth data acquisition. 
\section*{Laser characterization}
Here, we characterize the laser system performance by testing its outputs.
We start by investigating the polarization stability of critical outputs. 
This study leads to a discussion of the power stability, followed by frequency stability from the reference laser and an estimation of its frequency accuracy.
All tests in this section have been performed in a typical laboratory environment, i.e. temperature and humidity.
\subsection*{Polarization stability}
\begin{figure}[ht]
\centering
\includegraphics[width=1\linewidth]{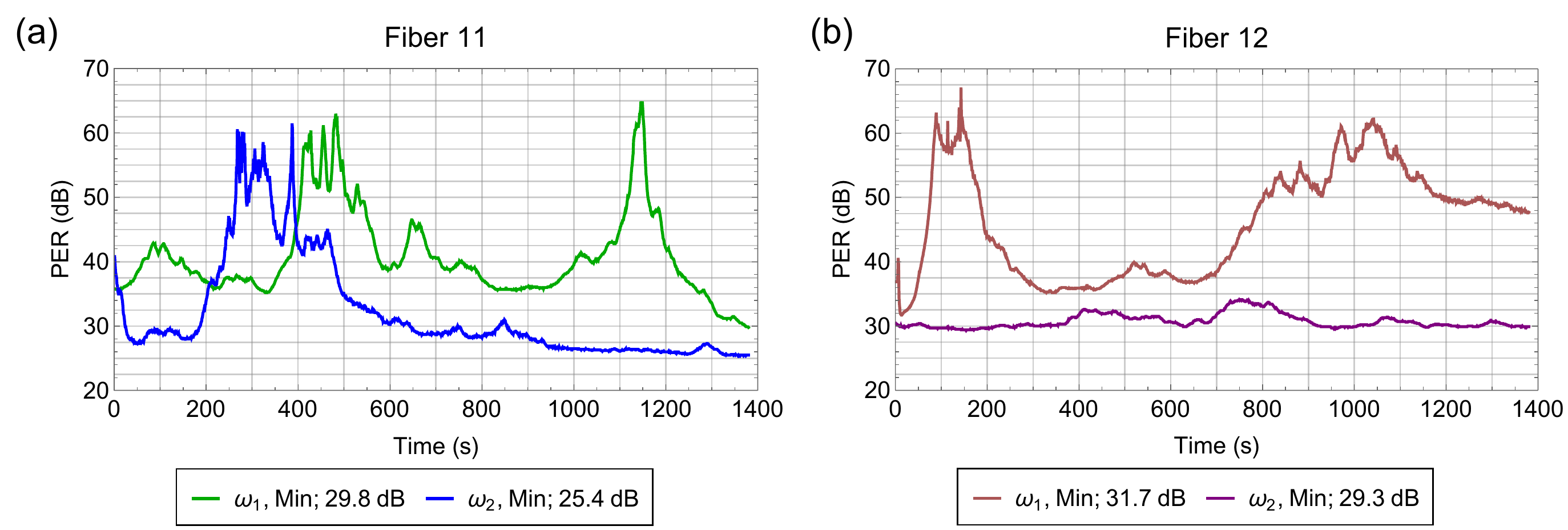}
\caption{PER of the Raman fibers. 
(a) PER of the two frequencies required to drive a Raman transition through the first Raman beam path - fiber (11).
Green is $\omega_{1}$, whose minimum PER is 29.8 dB, and blue is $\omega_{2}$, whose minimum PER is 25.4 dB.
(b) PER of the two frequencies required to drive a Raman transition through the second Raman beam path - fiber (12).
Brown is $\omega_{1}$, whose minimum PER is 31.7 dB, and purple is $\omega_{2}$, whose minimum PER is 29.3 dB.
\label{fig:polar}}
\end{figure}
Polarization stability is critical to our application in two ways: first, in our optical configuration, it directly translates to power instability, and so atom number fluctuation; second, polarization error of the Raman beams used in preparation and detection leads to inefficient pulses and unwanted light shifts.
To this end, we characterize the polarization stability by measuring the PER over time.
We report the minimum PER for each laser output over 3 hours in Table \ref{tab:one}, i.e. the worst PER obtained over the complete time series.
These tests are of the entire optical path required to get to the output of the fiber. 
In Figure~\ref{fig:polar}, we also present a time series of the PER of the Raman fibers (11) and (12) as an example as our most sensitive application.
We find that, for over three hours, the dominant fluctuation is thermal - the data sets shown are characteristic of the PER, shown up to 1380 s (for clarity). 
We find that all output fibers exhibit a minimum PER of 20 dB and, for the Raman outputs, we find a minimum PER of 25 dB.
\subsection*{Power stability}
The laser system must maintain both short- and long-term power stability. 
The power stability of the system is controlled by two independent feedback loops.
The first is a servo on the output power of the EDFA at 1560 nm.
The second is a servo for the PPLN crystal temperature - the doubling efficiency is typically a broad feature as a function of temperature and can be the source of power drift at long times.
We servo the temperature of the crystal; we have the option to servo the temperature of the crystal on the output power at 780 nm, from a monitor photodiode.
\par
\begin{figure}[ht]
\centering
\includegraphics[width=\linewidth]{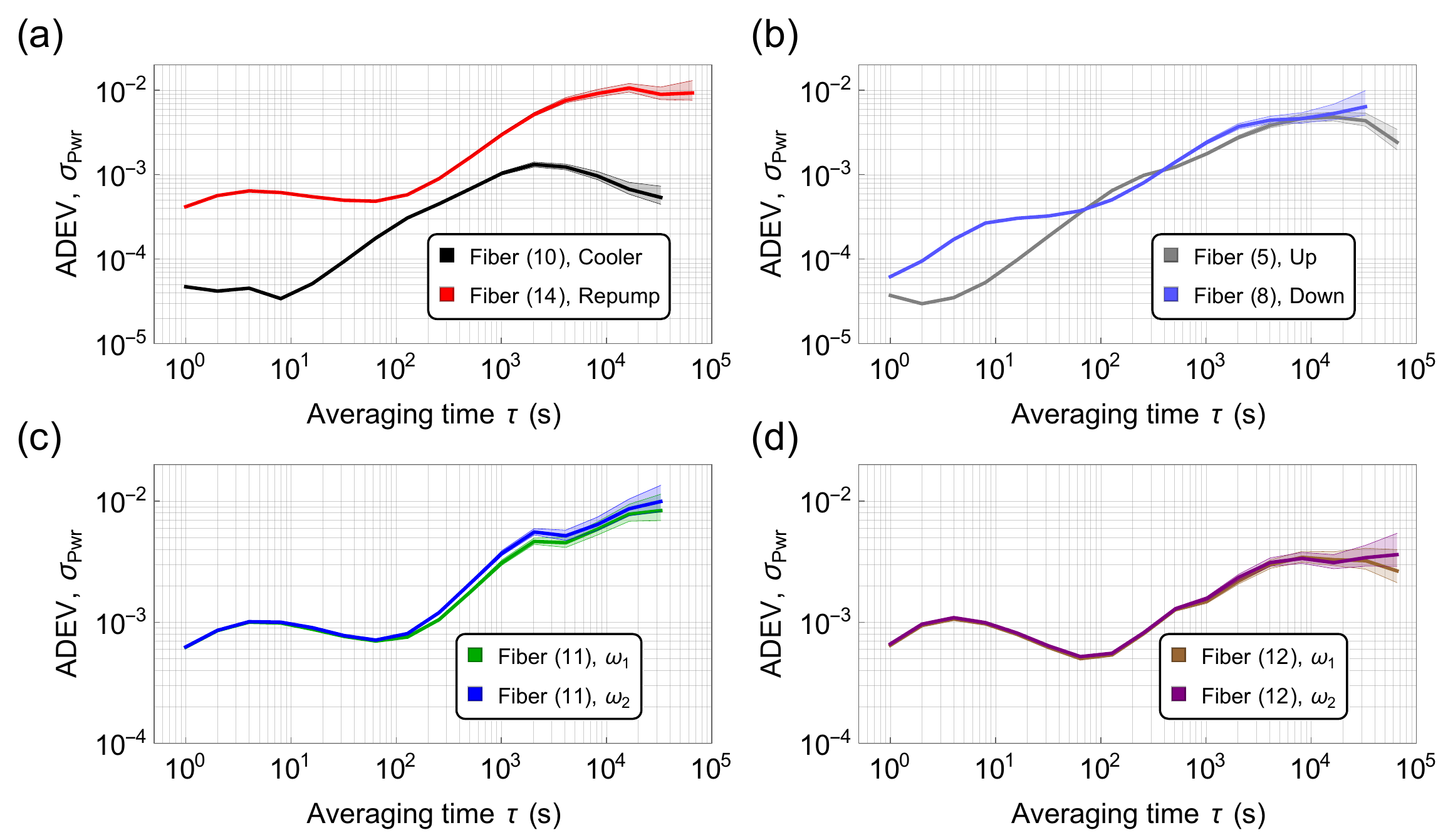}
\caption{Overlapping Allan deviation of laser power stability (normalized to output power). 
(a) Detection frequencies. 
Different total beam paths through the micro-optic modules lead to different power drifts. 
(b) 3D MOT "Up" versus "Down" balance. 
The beam paths are similar, which leads to similar low frequency drift.
(c) Power stability of both Raman frequencies in fiber (11). 
(d) Power stability of both Raman frequencies in fiber (12).
The shaded regions show the minimum and maximum extent.
\label{fig:pwrallan}}
\end{figure}

 We measured the laser power over time, sampling at 1 Hz, to observe any fluctuations and their characteristic time. 
We present the maximum output power and RMS of each output in Table~\ref{tab:one}.
We present some of these data, see Figure~\ref{fig:pwrallan}, for critical outputs as overlapping Allan deviations, relative to the output power level.
%
%
\par In (a) Figure \ref{fig:pwrallan}, we compare the power stability of the detection frequencies - the light from the cooling laser and the repump laser used to detect atoms.
In (b), we compare a fiber from the trio of 3D MOT up beams to a fiber from the trio of 3D MOT down.
In (c) and (d) of Figure~\ref{fig:pwrallan}, we compare the two laser frequencies that comprise the light in the fibers (11) and (12) - the Raman fibers. 
These power drifts are small, being a few parts in $10^{-3}$ on short timescales ($< 100$s) and $\simeq 10^{-2}$ on longer timescales ($10^{4}$ to $10^{5}$ s), which satisfies our requirements - this performance is comparable to and in some cases better than contemporary systems \cite{WANG201682, THERON2017152, caldani2019prototype}. 
If this performance needs to be improved, sufficient access to the AOMs is given to operate each AOM in an intensity noise-eater configuration.
\par
\begin{figure}[ht]
\centering
\includegraphics[width=0.5\linewidth]{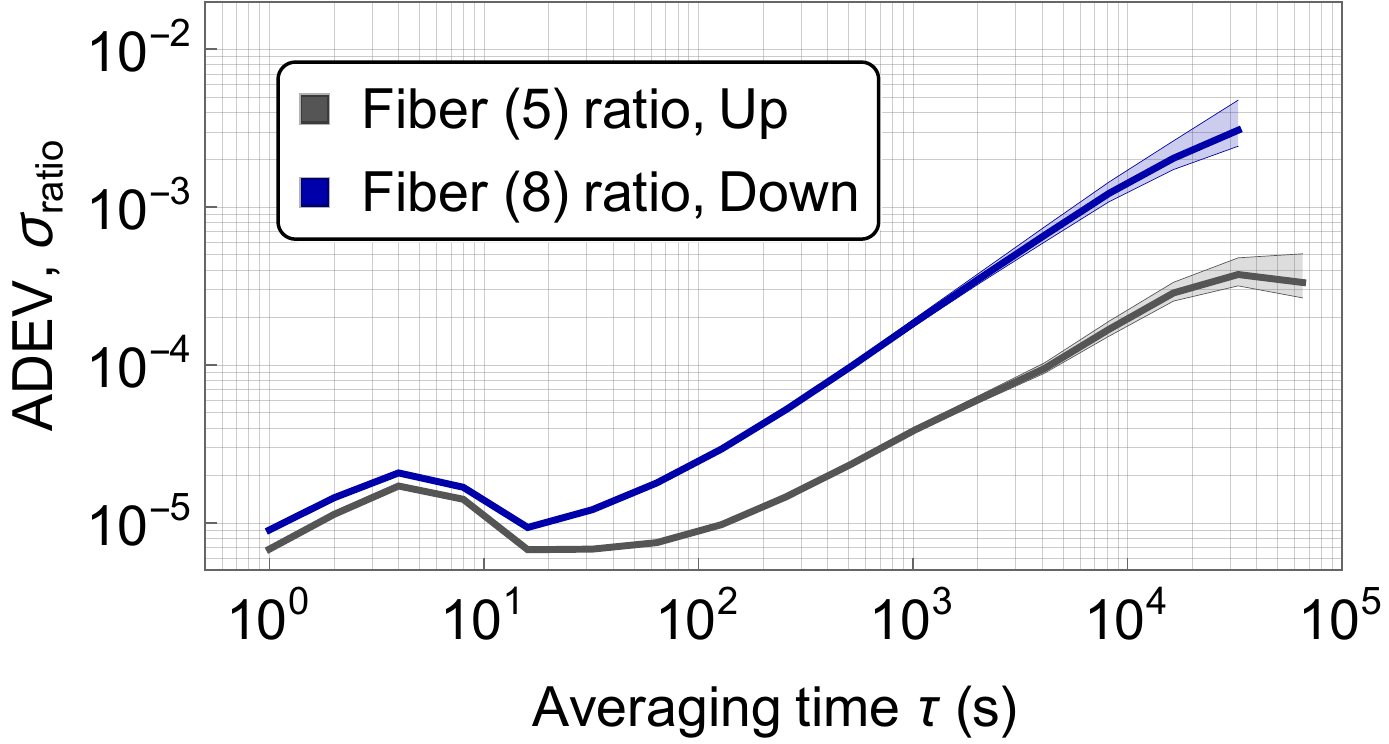}
\caption{Overlapping Allan deviation of laser power stability as a ratio to total fiber output in a micro-optic module (normalized to total output power). 
Shown for the "Up" and "Down" fibers, we observe similar behavior at short times but differing slopes and relative drift at long times. 
The shaded regions show the minimum and maximum extent.
\label{fig:updownratio}}
\end{figure}

 In Figure~\ref{fig:updownratio}, we show the same output of (b) in Figure~\ref{fig:pwrallan}, but as a ratio to the output of all fibers from the same micro-optic module.
This is a test of the stability through the micro-optic modules.
This is also an accurate estimate of the best performance an intensity noise-eater could obtain, as the power fluctuations presented here are no longer common mode. 
We find that power drifts between the outputs of the splitters are of the order of $\simeq 10^{-5}$ at short times ($\simeq 70$ s) and around $\simeq 10^{-3}$ at $10^{4}$ to $10^{5}$ s.
\subsection*{Frequency stability and estimated accuracy}
\begin{figure}[ht]
\centering
\includegraphics[width=.9\linewidth]{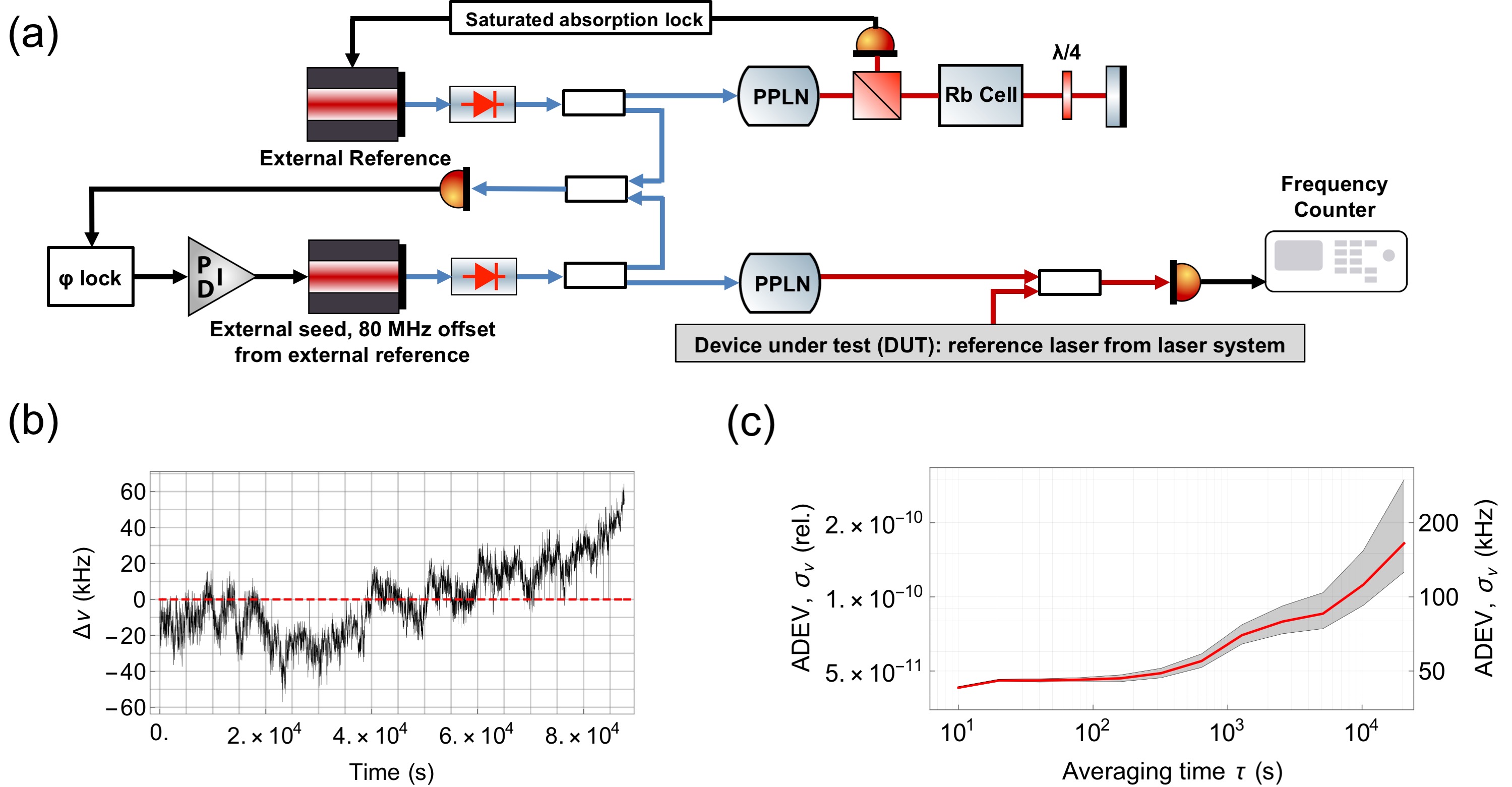}
\caption{Schematic and test of the reference seed frequency stability test versus an external reference laser.
(a) We compare our reference seed laser with a separate, external reference laser, pictured here, through an offset beatnote.
This external reference laser is frequency-stabilized to the same transition; an external seed is 80 MHz detuned and frequency-stabilized to this external reference. 
We then compare our reference seed signal, the DUT input, to this external seed laser.
(b) Raw data of the frequency drift with respect to the set point over time. Maximum difference seen in 25 hours is 64.5 kHz.
(c) Overlapping Allan deviation of the drift, with respect to the optical frequency. 
The shaded regions show the minimum and maximum extent. 
\label{fig:freqstabdata}}
\end{figure}

The laser system has an optical reference frequency, the reference laser in Figure~\ref{fig:laserspewpew}, using lock-in detection to the largest spectroscopic dip in a Rubidium spectrum with a natural abundance of the isotopes - the 3-4 crossover in $^{85}$Rb, $\ket{F=3}\rightarrow \ket{F=3}$ and $\ket{F=3}\rightarrow \ket{F=4}$ on the D$_2$ line. 
To this end, we split the output from the reference laser in two, at 1560 nm. 
The first path is frequency doubled to 780 nm and injected into a saturated absorption spectrometer \cite{Letokhov1976}, which in turn feeds a lock-in amplifier to servo the laser temperature and current. 
The second path is split between two mixers with two seed lasers for the creation of beatnotes.
Closing this servo loop creates a frequency reference with low drift and reduced environmental dependence.
We use a heterodyne technique to measure the absolute frequency stability of this reference signal using a beatnote with an external reference laser, shown in (a) of Figure \ref{fig:freqstabdata}. 
This external signal is obtained by first using a seed frequency-stabilized to the same dip in the Rubidium spectrum and then using a seed laser that is frequency stabilized to this external reference laser such that the beatnote between the seed and the reference laser is close to 80 MHz.
We make a beatnote between the reference laser in the laser system and the 80 MHz offset seed laser from the external reference system. 
We monitor the frequency difference of this beatnote and an 80 MHz reference signal using a frequency counter; the related data are presented in Figure~\ref{fig:freqstabdata}.
In (b) of Figure \ref{fig:freqstabdata}, we show a 25 hour test of this frequency difference sampled once every ten seconds.
We observe a minor trend; a low frequency liftoff at times $> 4 \times 10^{4}$ s.
In (c) of Figure~\ref{fig:freqstabdata}, we show the overlapping Allan deviation of these data, relative to the optical frequency. 
We find the frequency stability to be better than $5\times10^{-11}$ at 10 s and better than a few parts in $10^{-10}$ at $2 \times 10^{4}$ s.
Double Y-axes shows the ADEV relative to the frequency of the transition (384.2304844685(62) THz \cite{steck2001rubidium}) and relative to the setpoint (kHz).
The frequency reference for the 80 MHz offset is an ovenized quartz oscillator without GPS disciplining; this may be responsible for the drift at long times.
This is typical performance \cite{Masuda:07,Zhang_2012,leveque_2014,Matthey:15} from the saturated absorption spectrometer and its lock-in amplifier.
\section*{Tests using an atomic source of the MIGA antenna}
\begin{figure}[h!]
\centering
\includegraphics[width=1\linewidth]{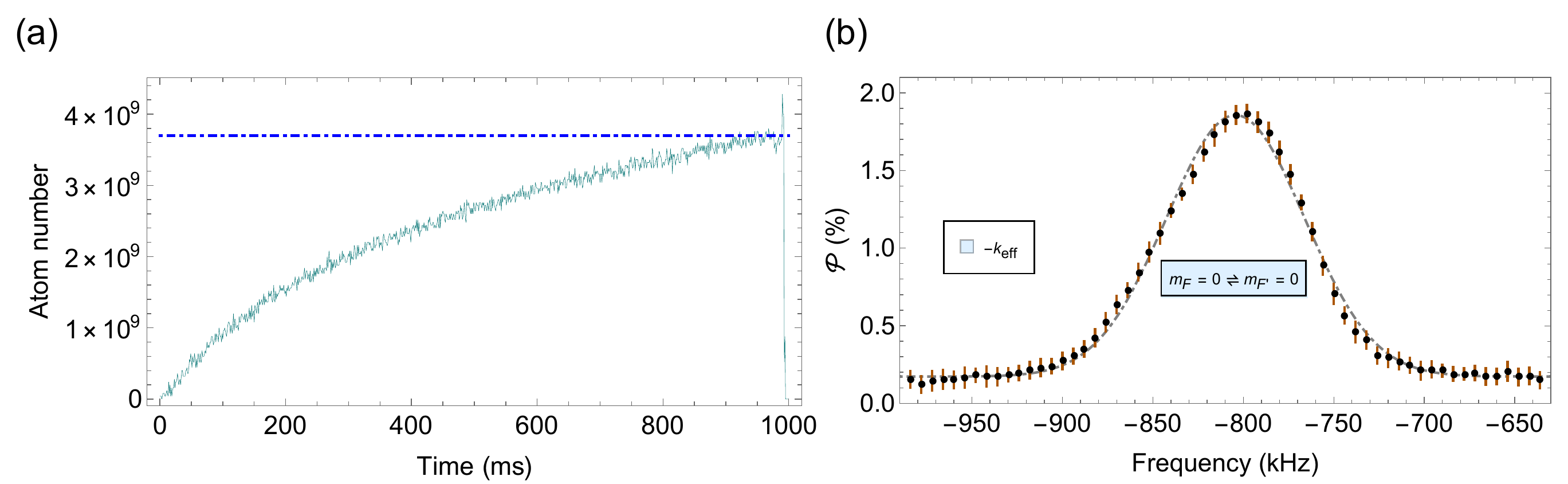}
\caption{3D MOT loading rate and temperature determination.
(a) Loading rate of the 3D MOT. At one second, we measure $3.7 \times 10^{9}$ atoms - the initial loading rate is $8.0 \times 10^{9}$ atoms.
(b) Temperature determination after launching and cooling. 
We fit a Gaussian model to the distribution - the gray dot-dashed curve is the fit. 
Each data point, in black, is the mean of ten measurements and the error bar, in orange, is the standard deviation of those ten measurements. 
We measure a temperature of 2.3 $\mu$K along the axis of the Raman beam, with no velocity preselection, using a 94 $\mu$s $\pi$-pulse.
This feature is the $\Delta m_{F}=0$, -$k_{\text{eff}}$ Doppler sensitive transition, which appears in Figure~\ref{fig:spectroscopy}.
\label{fig:atomfinal}}
\end{figure}
\begin{figure}[ht]
\centering
\includegraphics[width=1\linewidth]{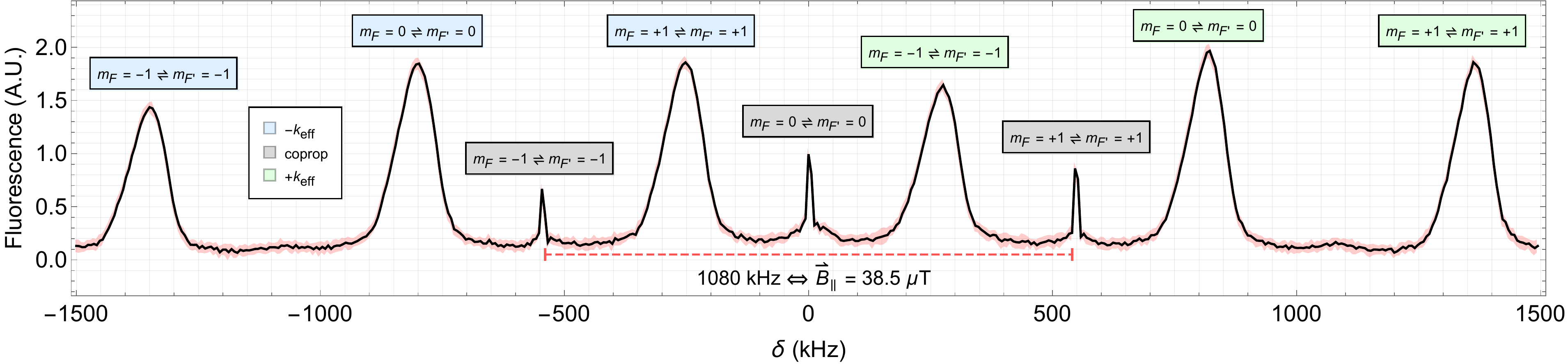}
\caption{Counter-propagating Raman spectroscopy - after cooling and launching atoms in the $F = 2$ manifold, with an applied magnetic field.
We scan the 3 MHz around the hyperfine splitting (6.834~GHz) symmetrically.
We use a $\pi$-pulse with characteristic time 94 $\mu$s.
We observe multiple features: the magnetic sublevels have their degeneracy lifted by 38.5~$\mu$T field, applied along the direction of the interrogation laser, the co-propagating transitions (labeled in gray) are visible due to slight polarization error on our injection mirror, and the Doppler-sensitive transitions ($\pm k_{\text{eff}}$ (green and red, respectively).
These data consist of 5000 points, with each point an average of 10 shots. 
The red shaded region is the minimum and maximum range of value.
\label{fig:spectroscopy}}
\end{figure}

To conclude our characterization of the laser system, we attach it to an atomic source - an exact model of the cold atom source that is used on the MIGA project \cite{Canuel2018}.
In brief, a horizontal 2D MOT loads a six-beam 3D MOT.
The atoms are launched, cooled, and then pass through a preparation region where the atoms undergo a velocity preselection along the horizontal direction parallel to the interrogation laser; during this they are also loaded into a magnetically insensitive substrate of ground states. 
They enter the interferometer interrogation region; afterwards, they fall back into this preparation region to be detected via fluorescence.
Details of the design and performance of the atomic source are not the subject of this paper.
Here, we present only the atomic measurements required to ascertain that the laser system meets the MIGA project requirements; more details on the atomic source are available\cite{spie, lefevre2017, remistuff, lefevre:tel-02163370}.
We measure a loading rate in excess of $3.7 \times 10^{9}~ \text{atoms}/\text{s}$ from the 2D MOT to the 3D MOT, see (a) of Figure~\ref{fig:atomfinal}.
After loading, we launch the atoms vertically at 4 m/s on a ballistic trajectory to reach an apogee of 85 cm above the 3D MOT - for the following tests we load the atoms into the $F = 2$ manifold where they populate the five Zeeman sublevels almost equally.
We use red-detuned Sisyphus cooling after the launch to bring the atomic ensemble's 3D kinetic temperature down.
In (b) of Figure~\ref{fig:atomfinal}, we show a temperature measurement of the ensemble - we use a long, velocity-selective Raman transition in a counter-propagating configuration to observe the Doppler-sensitive $\Delta m_{F}=0$, $\ket{F,m_{\text{F}}} \rightarrow \ket{2,0} \leftrightarrow \ket{1,0}$ transition, measuring a temperature of 2.3 $\mu$K along the axis of the beam.
Following this examination, we perform counter-propagating Raman spectroscopy, scanning a larger frequency range to observe co-propagating and Doppler-sensitive transitions, as well as observe the Zeeman splitting of all features, see Figure~\ref{fig:spectroscopy}. 
For the data of this last figure, we load the 3D~MOT from the 2D~MOT for 100~ms.  
The Raman lasers have a detuning $\Delta = -1.36$~GHz, see (b) in Figure~\ref{fig:laserimage}.
We scan $\delta$,  with respect to the hyperfine splitting of 6.834 GHz. 
We minimize the light shift by setting an intensity ratio of Raman 2/Raman 1 = 1.8 \cite{Peters2001}.
We use a constant intensity $\pi$-pulse, that is square in time, with a duration of 94~$\mu$s - this is in a Gaussian beam with a $1/e^{2}$ waist of 19.5~mm using low power ($\sim 20$ mW).  
We use a counter-propagating configuration - our frequencies come over the slow axis of the fiber ($\text{lin}\parallel\text{lin}$ input polarization) and are rotated by a $\lambda/4$ wave-plate at the retro-reflection mirror; this allows us to see both $\pm k_{\text{eff}}$ Doppler features.
The beams are tilted down with gravity by 5 degrees to observe the Doppler sensitive transitions.
We apply a nominal 40 $\mu$T magnetic field parallel to the beam path.
Without clearing any of the Zeeman sublevels, we see features corresponding to all possible $\Delta m_{F} = 0$ transitions.
We expect the atoms to be equally distributed among the Zeeman sublevels of the $F = 2$ state, with variations in this distribution due to non-adiabatic turn-off of the Sisyphus process in the polarization gradient cooling, slight power differences in the 3D MOT beams that lead to occurrences of optical pumping, and/or inhomogeneous magnetic fields and gradients through the region of cooling.
We find all expected properties, and that these spectroscopic features display a slight inhomogeneity in the normalized height of the features.
Using these data, we determine the applied magnetic field empirically to be 38.5 $\mu$T.
We do not find unexplained scalar, vector, or tensor light shifts.
Having observed Raman spectroscopy of the $F = 2$ manifold, we conclude that we can manipulate the atoms using Raman transitions with this laser system.
We do not use this laser system for atom interferometry with Raman transitions.
The MIGA project creates individual atom interferometers through Bragg diffraction, which is generated separately from this laser system. 
While not intended to interrogate an atom interferometer, this laser system has demonstrated the ability to make interference fringes \cite{lefevre:tel-02163370}.
\par
\begin{figure}[h!]
\centering
\includegraphics[width=1\linewidth]{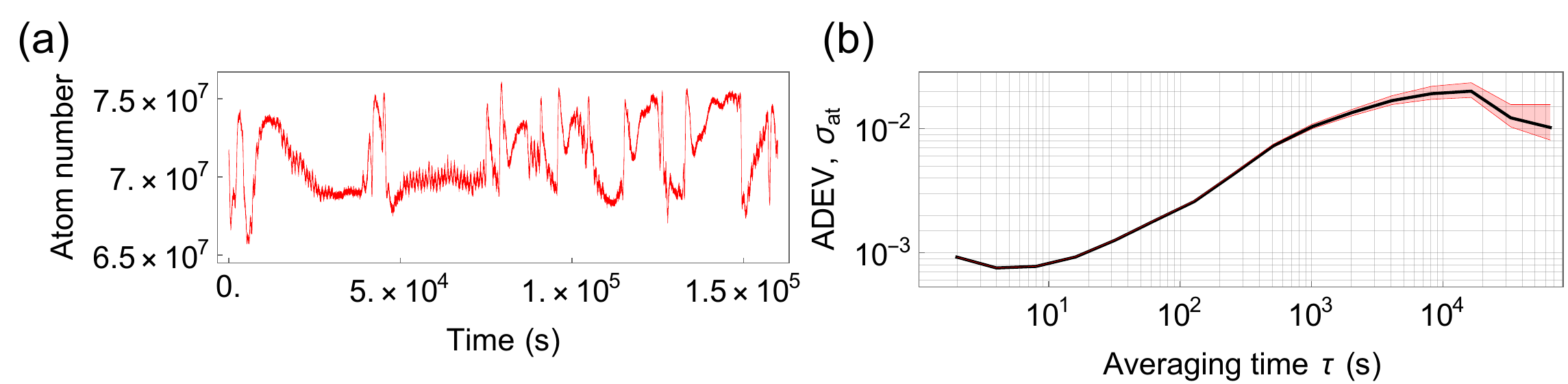}
\caption{Atom number stability.
(a) Atom number fluctuations over time.
(b) Overlapping Allan deviation of the number of atoms, normalized to the mean number of atoms. 
The red shaded region and thin red lines show the minimum and maximum extent. 
We observe a turning point just before $2\times10^{4}$ s. 
\label{fig:atomnb}}
\end{figure}

 Following this test, we measure the atom number stability as a function of time. 
We perform this test for over 4 days, see (a) of Figure~\ref{fig:atomnb}.
We analyze the fluctuations and present the overlapping Allan deviation normalized to the mean number of atoms, with the minimum and maximum extent displayed. 
These data were taken using the same sequence as the spectroscopy test.
With these findings, we conclude that the laser successfully integrates with our prototype and meets the application requirements for the cooling, trapping, manipulation, and detection of $^{87}$Rb. 
\section*{Conclusion}
We have designed and tested a laser system intended for the use of quantum technologies in a research infrastructure operating in a challenging environment.
Here, we present tests of the polarization, power, and frequency stability for this device over a long time.
Owing to the architecture and use of dedicated micro-optic modules, the system exhibits performances that meet the stringent requirements of the MIGA project. 
Using a prototype of the MIGA atomic source, we show that the system can cool, trap, manipulate, and detect $^{87}$Rb atoms. 
All tests reported here were carried out in a laboratory environment, for the purpose of a benchmark; the laser system has a thermal management system to create a mobile laboratory environment, ensuring its operation in varying conditions in terms of temperature and humidity. 
This will enable the system to operate within the underground LSBB facility.
\par Owing to the use of the industrial standard of fiber C-band telecommunications equipment, this system is reliable, flexible, and suitable for mass-production: there have been a total of five laser systems of this type produced for the MIGA project, all of which show comparable performance thanks to these technological choices.
Developing an industrial laser system that meets the requirements of cold atom physics is critical to the success of large scale atom interferometer projects such as MIGA \cite{Canuel2018}, ZAIGA \cite{Zhan2019}, ELGAR \cite{canuel2019elgar}, VLBAI \cite{Hartwig2015}, MAGIS/AION \cite{coleman2018magis100, magiswebsite, aionwebsite} and other national initiatives for atom interferometry presently under construction and consideration.
Furthermore, this laser system is an industrial asset for the production and operation of $^{87}$Rb quantum sensors: this device is easily miniaturized from this facility-scale form factor and, as such, is featured in mobile gravimeters, atomic clocks, and magnetometers.
\section*{Data availability}
The data presented and analyzed in this study are available from the corresponding author upon reasonable request.
%
%
\bibliography{main.bbl}

\begin{thebibliography}{10}
\bibitem{remistuff}
\bibinfo{note}{{A manuscript on the performance and operating parameters of the
  MIGA atomic source is in production for submission to a special topic of EPJD
  on quantum technologies for gravitational physics. Citation to be added.}}
\end{thebibliography}
\section*{Acknowledgements}
This work was realized with the financial support of the French State through the ``Agence Nationale de la Recherche'' (ANR) within the framework of the ``Investissement d'Avenir'' programs Equipex MIGA (ANR-11-EQPX-0028) and IdEx Bordeaux - LAPHIA (ANR-10-IDEX-03-02). This work was also supported by the r{\'e}gion d'Aquitaine (project IASIG-3D and USOFF). We also acknowledge support from the CPER LSBB2020 project; funded by the ``r{\'e}gion PACA'', the ``d{\'e}partement du Vaucluse'', 
and the ``FEDER PA0000321 programmation 2014-2020''. We acknowledge the financial support from Ville de Paris (project HSENS-MWGRAV) and Agence Nationale pour la Recherche (project PIMAI, ANR-18-CE47-0002-01). G.L. thanks DGA for financial support. X.Z. thanks the China Scholarships Council (N$^\mathrm{o}$ 201806010364) program for financial support. J.J. thanks ``Association Nationale de la Recherche et de la Technologie'' for financial support (N$^\mathrm{o}$ 2018/1565). 
\subsection*{Contributions}
The conception of the laser architecture and primary tests were done by B.B., A.L., and P.B.
The laser system was designed by A.B., G.S., A.L., B.D., P.B., and B.C.
The system was built and tested by G.S., J.S., and B.D.
The manuscript was written by D.S. and B.C. The laser data were analysed by D.S., J.J., G.S., and B.C.  
The tests of the laser system on the MIGA cold atom source were carried out by D.S., J.J., G.L., X.Z., A.B., M.P., Q.B., R.G., A.L., P.B., and B.C. The atomic data were analysed by D.S., J.J., and B.C.
All the authors contributed to manuscript preparation, as well as having read and approved the final manuscript.
\subsection*{Competing interests}
Muquans is a startup company, created in 2011, which exploits results of LP2N and SYRTE. 
B.D., A.L., and P.B. are members of the strategic committee of Muquans and own stock in the company. 
B.D. is the C.E.O. of Muquans; J.J., G.S., and J.S. are employed by the company. 
A.L. and P.B. receive compensation as scientific advisors for Muquans. 
Others declare no potential conflict of interest.
\subsection*{Corresponding author}
Please send correspondence to benjamin.canuel@institutoptique.fr
\end{document}